\newcommand{\diracslash}[1]{#1\llap{/\kern2pt}}

\newcommand{\be}{\begin{equation}}
\newcommand{\ee}{\end{equation}}
\newcommand{\bea}{\begin{eqnarray}}\index{\footnote{}}
\newcommand{\eea}{\end{eqnarray}}
\newcommand{\ba}[1]{\begin{array}{#1}}
\newcommand{\ea}{\end{array}}

\documentclass[onecolumn,secnumarabic,amssymb,nobibnotes,nofootinbib,showpacs]{revtex4}
\usepackage{graphicx}
\usepackage{mathtools}
\usepackage{amssymb}
\usepackage{mathrsfs}
\usepackage{color}
\usepackage{rotating}
\usepackage{hyperref}
\usepackage{cleveref}
\begin{document}
\setlength{\topmargin}{-0.5in}
\title{A model study on superfluidity of a unitary Fermi gas of atoms interacting with a finite-ranged potential}
\author{Subhanka Mal and Bimalendu Deb}
\affiliation{School of Physical Sciences, Indian Association for the Cultivation of Science, Jadavpur, Kolkata 700032, India.}
\begin{abstract}
We calculate Bardeen-Cooper-Schrieffer (BCS) state of a unitary Fermi gas of atoms interacting with the finite-ranged Jost-Kohn potential which has been recently shown to account for the resonant interactions [2019 {\rm J. Phys. B: At. Mol. Opt. Phys.} {\bf 52}, 165004]. Using exact scattering solution of the potential, we derive two-body ${\mathbf T}$-matrix element which is employed to construct the BCS Hamiltonian in momentum space. We present results on the energy- and range-dependence of the pairing gap and superfluid density and the range-dependence of the chemical potential for a wide variation of the scattering length including the unitary regime. In the zero range limit our calculated gap at the Fermi energy is found to be nearly equal to that calculated in mean-field theory with contact potential. The mean gap averaged over the full width at half maximum of the gap function in the zero range and unitary limits is found to be $0.42 E_F$ which is quite close to the recent result of the quantum Monte Carlo simulation [2018 {\rm Phys. Rev.A} {\bf 97}, 013601]. The chemical potential in the zero range limit also agrees well with that for the contact potential.
\end{abstract}
\maketitle
\section{Introduction}\label{sec.1}

The experimental advancement in many-body quantum physics \cite{Bloch_RMP:2008} with trapped ultracold atoms over the last two decades 
has opened new perspectives in both bosonic and fermionic superfluidity. In particular, the studies of superfluidity with a Fermi gas of atoms  
have implications in understanding the superconductivity of condensed matter systems, superfluidity of He-3 liquid and superconductivity or superfluidity of nuclear matter of astrophysical origin.  It is expected that quantum simulations of various types of fermionic superfluidity using trapped atomic gases with control over atom-atom interactions, trap dimensionality  and atomic density will provide new insight into electronic superconductivity and nuclear superfluidity.  Superfludity and superconductivity occur due to the spontaneous breaking of continuous symmetry resulting in the appearance of long wave-length Goldstone modes which are also known as  Bogoliubov-Anderson modes. While Goldstone modes in charged superconductors become ill-defined due to 
Coloumb interactions, such modes in neutral superfluidity are expected to be well-defined and detectable. Bogoliubov-Anderson modes have been recently measured in a superfluid Fermi gas of atoms \cite{Hoinka_Nat_Phy:2017,Kuhn_PRL:2020}. Thus the charge neutrality of atomic superfluidity offers some advantages in exploring fundamental aspects of quantum many-body physics. On the other hand, the aspects of superconductivity that are intrinsically connected with charge  or gauge fields are not available with the neutral systems. Nevertheless, in recent times, artificial gauge fields for neutral atomic gases are created by optical manipulations of  atomic internal as well as center-of-mass degrees-of-freedom \cite{Dalibard_RMP:2011}. 

Bardeen-Cooper-Schrieffer (BCS) \cite{BCS_PR:1957} type or standard superconductivity in electrons results from the phonon-mediated electron-electron attraction near the Fermi surface. This attractive interaction overcomes the Coulomb repulsion between the electrons at low energy, resulting in the formation of Cooper pairs which then constitute the many-body BCS state.   Fermi superfluidity in trapped atomic gases arises when the atom-atom  interaction is tuned towards attractive side by a magnetic Feshbach resonance \cite{Chin_RMP:2010}. In general, the ultracold atomic gases that are currently being experimentally studied are sufficiently dilute such that the range of interaction between the atoms is very small compared to the interatomic separation. Therefore,  the range is generally neglected and the interaction is approximated as a contact potential expressed in terms of the $s$-wave scattering length $a_s$. A magnetic field is used to tune $a_s$ across the Feshbach resonance. A degenerate Fermi gas 
is transformed into BCS state by tuning $a_s$ towards small negative value for which the effective atom-atom interaction becomes attractive. Near the resonance where $a_s$ is large negative, the gas becomes strongly interacting leading to resonance superfluidity \cite{Kokkelmans_PRA:2002} that is characterized by the unitarity of the scattering $S$-matrix. 
On the positive side of $a_s$ across the resonance, the Cooper pairs may be transformed into bosonic diatomic molecules which can condense into a Bose-Einstein condensate. Such  unitary regime and crossover physics may be available with  color superconductivity in quantum chromodynamics and neutral superfludity of neutron stars.  

Given the current state of the art in the experimental cold atom physics, an experimental attempt towards simulating neutron superfludity appears to be a difficult task.   However, the recent successful observations of BCS-BEC crossover   
open prospect for simulating  nuclear physics  \cite{Randeria_PRL:1989,Micnas_RMP:1990,Randeria_PRL:1992,Drechsler_AnnPh:1992,Casas_PRB:1994}. 
The effective range of interaction  plays an important role in many-body quantum dynamics when it becomes comparable to inter-particle separation \cite{Horikoshi_PRX:2017,Ohashi_PRA:2018,Horikoshi_IJMPE:2019,Yin_PRL:2019}.
In recent times, there has been considerable research interests to simulate dilute nuclear matter \cite{Baker_PRC:1999} such as proto-neutron stars \cite{Heckel_PRC:2009,Baym_RPP:2018}.  From a theoretical  point of view, most of the works have been carried out by numerical experiments i.e., by Monte-Carlo simulations \cite{Navon_Sci:2010} or using density functional theory \cite{Frobes_PRL:2011}. The effects of the finite range of interaction on universal equation of state (EOS) of the system have been theoretically studied using several model finite-ranged two-body interaction potentials \cite{Frobes_PRA:2012,Santiago_PLA:2013,Neri_PS:2020}. The thermodynamic behaviour of a Fermi gas has been studied as a function of the parameters of a model finite-ranged potential of exponential type in the BCS-BEC crossover regime \cite{Santiago_PLA:2013}. Two different class of finite-ranged potentials - one purely attractive such as the square well, the exponential and the Yukawa type potential, and the other having both attractive and repulsive character such as Van der Waals and dipolar potentials have been considered to explore how the depth and the spatial range of these potentials affect the pairing and molecule formation along the crossover \cite{Neri_PS:2020}. Spin fluctuations of a strongly interacting Fermi gas across BCS-BEC crossover have been demonstrated by  speckle imaging \cite{Sanner_PRL:2011}. Momentum resolved photo-emission  spectroscopy has been used to observe many-body pairing of a two dimensional trapped unitary Fermi gas above the transition temperature \cite{Feld_Nat:2011}. Controllable quasi-2D system across crossover has been observed with results that are beyond  the mean-field level \cite{Makhalov_PRL:2014}. Quantized vortex ring is also observed in a unitary Fermi gas \cite{Yefsah_Nat:2013,Bulgac_PRL:2014}.

The purpose of this paper is to explore the effects of the effective range 
of interaction on the superfluidity of a unitary Fermi gas. Towards this end, we resort to the finite-range Jost-Kohn (JK) model potential \cite{Mal_JPB1:2019} which has been recently shown to account for the unitary regime. The use of this model  interaction potential allows us to study the finite range effects and energy dependence of superfluid gap and density over a wide range of $a_s$ including the resonance limit.  We use exact scattering solution of the JK potential in order to calculate momentum-dependent two-body ${\mathbf T}$-matrix element for the BCS Hamiltonian.  
We present results on the energy-dependence of the superlfuid pairing gap over the entire range of energy for different values of $a_s$ and range. The gap is found to exhibit strong energy dependence in the unitary regime. We also present results on the effects of the range and $a_s$ on the superfluid density distribution. Our results in the zero-range limit  qualitatively agree well with those for the contact interaction potential. We find that a mean gap averaged over the energy within the full width at half maximum of the gap function in the zero energy limit of our model interaction potential agrees quite well with the value calculated recently by quantum Monte Carlo simulation \cite{Ohashi_PRA:2018}.  
 
 The paper is organised in the following way.  In section \ref{sec.2} we discuss the standard BCS theoretical methods, introduce the Jost-Kohn model interaction potential and its exact scattering solution. In section \ref{sec.3} we present and discuss our results showing the effects of finite range and comparing them with those for contact interaction. We conclude and give an outlook of the work in section \ref{sec.4}.

% Ultracold atoms has been studied widely owing its experimental realization \cite{Jin_PRL:2004,Ketterle_PRL:2004,Thomas_PRL:2004,Bartenstein_PRL:2004} of trapped atomic Fermi gas and condensate of atomic Bose gas\cite{Ketterle:1995,Cornell:1995}. The observation of crossover from superfluid Fermi gas to condensate of bosonic molecules by precise tuning of interaction via so called Feshbach resonance has become a pathway for experimentalists to simulate condensed matter systems \cite{Feynman_IJTP:1982,Bloch_Nat:2012}. From realization of degenerate Fermi gas \cite{Jin_Sci:1999} to Superfluid-to-Mott insulator transition \cite{Bloch_Nat:2002,Esslinger_PRL:2004} in an optical lattice is shown to be achievable by tuning the interaction between atoms. The high tunability also enabled researchers to study quantum many-body problems and most importantly quantum simulation \cite{Bloch_Nat_Phys:2012}. Superfluid gap has been accurately determined by means of quasiparticle spectroscopy with radio frequency \cite{
%Schirotzek_PRL:2008} for a strongly interacting imbalanced Fermi gas.

\section{Theoretical methods}\label{sec.2}
According to the Bardeen-Cooper-Schrieffer (BCS) theory \cite{BCS_PR:1957} of  low temperature superconductivity,  an attractive interaction between fermions in a quantum degenerate Fermi system leads to the formation of Cooper pairs \cite{Cooper_PR:1956} and the instability of the Fermi surface. At a critical temperature, the system then undergoes transition to superconducting phase. In solid-state electronic systems, 
a pair of electrons near the Fermi surface interacts by means of lattice-assisted or phonon-mediated attractive interaction, resulting in the formation of $s$-wave Cooper-pairs  in spin-singlet state. The two electrons that form a Cooper-pair have equal and opposite lattice momenta. The Cooper-pair wave function is given by 
\bea
\psi_0({\bf r_1,r_2})=\sum_{k>k_F}g_{\bf k}e^{i{\bf k.}({\bf r_1-r_2})}
\eea
% Taking the symmetric spatial part,
% \bea
% \psi_0({\bf r_1,r_2})=\sum_{k.k_F}g_{\bf k}\cos{\bf k.(r_1-r_2)}(\uparrow_1\downarrow_2-\uparrow_2\downarrow_1)/\sqrt{2}
% \eea
where ${\bf r}_{1(2)}$ refers to the position vector of the electron 1 (2) and $g_{\bf k}$ is the wave function in momentum space. On substitution of the above relation into Schr\"{o}dinger equation, one obtains 
\bea
(E-2\epsilon_{\bf k})g_{\bf k} = \sum_{k'>k_F}g_{\bf k'}V_{ \bf kk'}
\label{eq2}
\eea
where $E$ is the eigen energy, $\epsilon_k=\hbar^2k^2/2\bar{m}$ ($\bar{m}$ is reduced mass) and  $V_{\bf k k'}$ is the interaction in the momentum space. In the weakly interacting regime, $V_{\bf k k'}$ is related to the real-space interaction potential $V_{int}( r)$ by  
\bea
V_{\bf k k'}=\frac{1}{(2 \pi)^3} \int V_{int}(r)e^{-i({\bf k-k'}).{\bf r}}d^3r
\label{eq3}
\eea
where ${\bf r} = {\bf r}_1 - {\bf r}_2$ is the relative position vector of the two fermions. Approximating the two-particle interaction by the 
 Fourier transform of $V_{int}(r)$ into the momentum space as given above is akin to first order Born approximation of scattering ${\mathbf T}$-matrix. 
In case of a conventional superconductor, $V_{int}$ is the phonon-mediated potential and is often assumed to be a space-independent contact 
potential. As a consequence, the momentum integration in Eq. (\ref{eq2}) diverges. This divergence is tackled by putting a cut-off as the upper limit in the energy or wave number. In case of low temperature standard superconductors, there exists a natural cut-off which is the Debye energy or Debye wave number. 

The BCS theory is applicable to superfluidity  of neutral Fermi liquids or gases, albeit with some modifications that are necessary due to the absence of any natural cut-off in  energy.  For neutral atomic Fermi gases, $V_{int}(r)$ may be, in general, of contact or finite-range or even  long-range type potential. If $V_{int}(r)$ is approximated as a contact potential represented by a delta function, $V_{\bf k k'}$  becomes a constant that is proportional to the $s$-wave scattering length $a_s$. Such approximation leads to logarithmic divergence in momentum  integration of gap equation. This divergence  is overcome by renormalising the interaction parameter or the coupling constant by subtracting the ultraviolet divergence part from the integrand \cite{Randeria_PRL:1989, Deb_PRA:2004}. However, for a short-range or finite-range interaction, such divergence problems will not arise and the gap equation or superfluid behavior will depend on the range of interactions apart from other system 
parameters.

Retaining only those interaction terms where two fermions interact with zero center-of-mass momentum, one can write down the reduced  Hamiltonian 
\bea
{\cal H}=\sum_{k,\sigma}\epsilon_k\hat{n}_{k\sigma}+\sum_{k,k'}V_{kk'}\hat{c}_{k\uparrow}^\dagger\hat{c}_{-k\downarrow}^\dagger\hat{c}_{-k'\downarrow}\hat{c}_{k'\uparrow}
\label{eq4}
\eea
 In  Hartree self consistent or generalised mean-field approach, the BCS ground state is given by generalised spin coherent state \cite{Montorsi_PRB:1997} 
\bea
|\psi_G\rangle=\prod_{k} \left (u_k+v_k\hat{c}_{k\uparrow}^\dagger \hat{c}_{-k\downarrow}^{\dagger} \right )|\phi_0\rangle
\label{eq5}
\eea
where $|v_k|^2$ is the probability of the pair-state $({\bf k},-{\bf k})$ being occupied and $|u_k|^2+|v_k|^2 = 1$. Here $|\phi_0\rangle$ is the vacuum state which is annihilated by $\hat{c}_k$. 
Introducing the number operator  $\hat{N}=\sum_k(\hat{c}_{k\uparrow}^\dagger\hat{c}_{k\uparrow}+\hat{c}_{k\downarrow}^\dagger\hat{c}_{k\downarrow})$, and a variational parameter 
$\theta_k$ such that $v_k=\cos\theta_k$ and $u_k=\sin\theta_k$, one can obtain the BCS ground state by using the variational principle \cite{deGennes:1999}
\bea
\delta\langle\psi_G|{\cal H}-\mu\hat{N}|\psi_G\rangle=0
\label{eq6}
\eea
Here $\mu$ is the chemical potential which conserves the total number of particles. $\cal H$ of equation (\ref{eq4}) can be represented in terms of pseudospin operators \cite{Montorsi_PRB:1997}. The value of $\langle\cal H\rangle$ is calculated by implementing equation (\ref{eq4}) and (\ref{eq5}).  Thus we have 
\bea
\langle{\cal H}-\mu\hat{N}\rangle=\sum_k\xi_k(1+\cos2\theta_k)+\frac{1}{4}\sum_{k,k'}V_{kk'}\sin2\theta_k\sin2\theta_{k'}\nonumber\\
\label{eq7}
\eea
where $\xi_k=(\epsilon_k-\mu)$. Taking the variation with respect to $\theta_k$ we have
\bea
\frac{\partial}{\partial\theta_k}\langle{\cal H}-\mu N\rangle=-2\xi_k\sin2\theta_k+\sum_{k'}V_{kk'}\cos2\theta_k\sin2\theta_{k'} =0\nonumber\\
\label{eq8}
\eea
which results in 
\bea 
\tan2\theta_k=\frac{1}{2\xi_k}\sum_{l}V_{kk'}\sin2\theta_{k'}
\label{eq9}
\eea 
Now, defining the  superfluid gap  $\Delta_k=-\sum_{k'}V_{kk'}u_{k'}v_{k'}=-\frac{1}{2}\sum_{k'}V_{kk'}\sin2\theta_{k'}$, the above equation can be expressed in the form of the well-known gap equation 
\begin{eqnarray}
\Delta_k=-\frac{1}{2}\sum_{k'}\frac{V_{kk'}\Delta_{k'}}{\sqrt{\Delta_{k'}^2+\xi_{k'}^2}}
\label{eq:gap}
\end{eqnarray}
with two probability amplitudes being $u_k=\frac{1}{\sqrt{2}}\left(1+\xi_k/\sqrt{\Delta_k^2+\xi_k^2}\right)^{\frac{1}{2}}$ and $v_k=\frac{1}{\sqrt{2}}\left(1-\xi_k/\sqrt{\Delta_k^2+\xi_k^2}\right)^{\frac{1}{2}}$.

\subsection{Finite-ranged model potential and its solution}\label{subsec.2.1}
To study the effects of the range of interaction, we use  JK interaction potential  
\begin{eqnarray}
  V_{int}(r)=-\frac{4\hbar^2}{\bar{m} r_0^2}\frac{\alpha\beta^2e^{-2\beta r/r_0}}{[\alpha+e^{-2\beta r/r_0}]^2}
  \label{eq:JK-}
\end{eqnarray}
where $r=|{\bf r_1-r_2}|$, $\alpha=\sqrt{1-2r_0/a_s}$, $\beta=1+\alpha$, ${\bf r_1}$ and ${\bf r_2}$ are the coordinates of two particles and the $s$-wave scattering length $a_s<0$. 

In order to access the strongly interacting regime with  a finite-ranged interaction, it is necessary to employ the exact scattering solution of the two-body problem as the first order Born approximation breaks down. In the limit of large $a_s$, the JK potential of eq.(\ref{eq:JK-})
takes the form Ps\"{o}chl-Teller potential which admits an exact solution 
\cite{Deb_IJMPB:2016,Flugge_Book:1998} given by
%\begin{widetext}
\bea
\hspace{-0.9in}
\psi_k(r)=\cosh^\lambda{(\kappa r)}\sinh{(\kappa r)} _2F_1\Bigg(\left(\frac{\lambda}{2}+\frac{ik}{2\kappa}+\frac{1}{2}\right),\left(\frac{\lambda}{2}-\frac{ik}{2\kappa}+\frac{1}{2}\right),\nonumber\\
\frac{3}{2};-\sinh^2{(\kappa r)}\Bigg)
\eea
%\end{widetext}
where $\kappa=2/r_0$, $\lambda=\frac{1}{2}\left(1-\sqrt{1+8/\alpha}\right)$ and $_2F_1(a,b,c;z)$ is the hypergeometric function.  This wave function has the asymptotic form 
\bea
\psi_k(r\rightarrow\infty)\sim\pm C\cos{(kr+\phi_0)}
\eea
where 
\begin{equation}
C=2\Gamma(3/2)\left|\frac{\Gamma(-ik/\kappa)\exp({ik\log{2}/\kappa})}{\Gamma(\frac{\lambda+1}{2}+i\frac{k}{2\kappa})\Gamma(1-\frac{\lambda}{2}-i\frac{k}{2\kappa})}\right|
\end{equation}
and
\bea
\phi_0={\rm arg}\left[\frac{\Gamma(-ik/\kappa)\exp({ik\log{2}/\kappa})}{\Gamma(\frac{\lambda+1}{2}+i\frac{k}{2\kappa})\Gamma(1-\frac{\lambda}{2}-i\frac{k}{2\kappa})}\right]
\eea 
Now, we normalize the wave function $\psi_k^{sct}(r)\rightarrow\psi_k(r)/C$ in a way so that the asymptotic form reduces to $\psi_k^{sct}(r\rightarrow\infty)=\sin(kr+\delta_0)$ where $\delta_0=\phi_0+\pi/2$ is the $s$-wave phase shift. The effective interaction $V_{kk'}$ in momentum space can be obtained from the exact on-the energy shell ${\mathbf T}$-matrix element. Explicitly, we have 
\bea
 V_q=2\pi\int j_0(qr)V_{int}(r)\frac{\psi_q^{sct}(r)}{qr}r^2dr
\eea
where $q=|\bf{k-k'}|$ and $j_0(qr)=\sin{(qr)}/qr$ is the spherical Bessel function of zeroth order. To extract $V_{kk'}$ from $V_q$ we write $q=\sqrt{k^2+k'^2-2kk'\cos\theta}$. To get the exclusive dependence on $k$ and $k'$ we must integrate $V_q$ over $\theta$ which then yields
\begin{eqnarray}
V_{kk'}=\int_{|k-k'|}^{(k+k')}\frac{V_qq}{kk'}dq \hspace{0.5in} k\neq0\neq k'\nonumber
\end{eqnarray}
For $k=0$, $V_{0k'}=V_q$ and for $k'=0$, $V_{k0}=V_q$. Also for $k=0=k'$ we have to take $V_{00}=V_q(q=0)$. From the numerical point of view we first generate a large set of discrete values of $q$ and corresponding $V_q$. By this procedure we obtain a two-dimensional array of $V_{kk'}$ as a function of $k$ and $k'$.
\begin{figure}[htbp]
\centering
 \includegraphics[height=1.7in, width=5.4in]{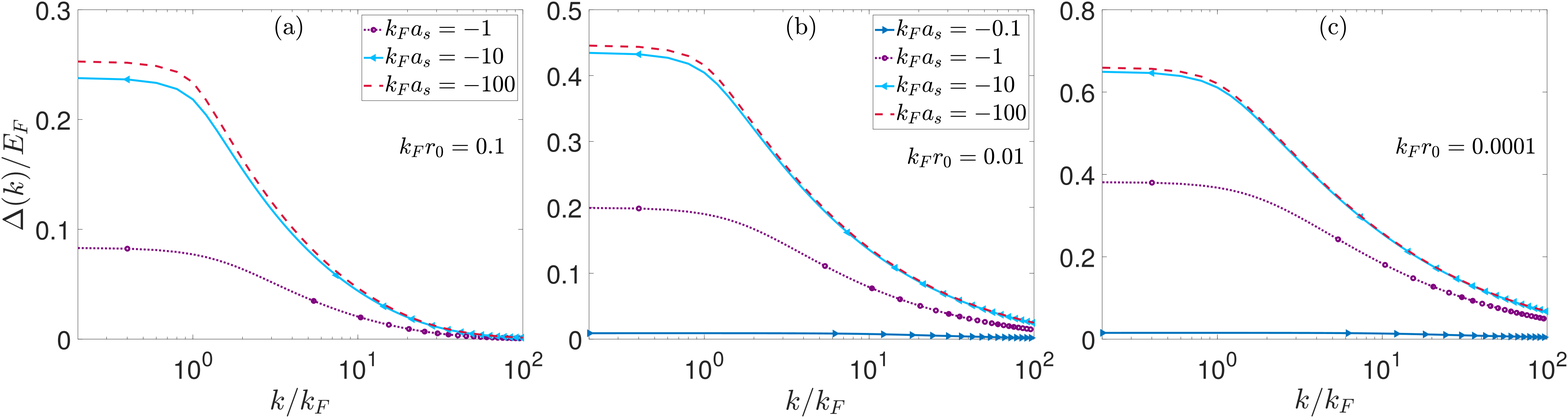}
 \centering
 \caption{The superfluid gap $\Delta(k) \equiv \Delta_k$ (in unit of Fermi energy $E_F$) as a function of wave number $k$ (in unit Fermi wave number) is shown for different scattering lengths $a_s$ and different values of the range $r_0$. The values of $a_s$ in (c) are same as in (b).}
 \label{fig.1}
\end{figure}

The JK potential is expressed in terms of two parameters, namely the $s$-wave scattering length and the effective range of a real two-body interaction potential. It is derived by inverse scattering method. So, our theoretical method is applicable to all the interacting Fermi systems that admit effective range expansion of the scaterring phase shift. So unlike various types of model potentials used in the current literature \cite{Frobes_PRA:2012,Santiago_PLA:2013,Neri_PS:2020}, the use of JK potential offers some advantage in numerical computation in that there is no need for the calculation of scattering length and the effective range which enter into the BCS Hamiltonian as the input parameters. But the most significant feature of the JK potential is that
it can capture the resonance effect or unitary regime of interaction for which it is necessary to use exact scattering solution of the potential. However, this potential can not describe the effects of actual spatial range of a realistic potential.

\subsection{k-dependent gap and superfluid density}\label{subsec.2.2}
Here we consider the superfluid gap as a function of momentum. By solving the gap equation in a self-consistent manner we calculate the $k$-dependent gap. The equation(\ref{eq:gap}) is rewritten in the continuous form
\bea
\Delta_k=-\frac{1}{2}\int\frac{V_{kk'}\Delta_{k'}}{\sqrt{\Delta_{k'}^2+(\epsilon_{k'}-\mu)^2}}\frac{d^3k'}{(2\pi)^3}
\label{eq:delk}
\eea
One point worth noticing here is that we need not take into account the renormalization of the interaction term unlike that in the standard gap equation with contact potential. However, in our case the interaction in $k$-space goes to zero as $k\rightarrow\infty$. This allows us to obtain convergent solutions of $\Delta_k$ by a numerical iteration process.

The $k$-dependent superfluid density $n_k$ or equivalently $n(E)$ is given by $n(E) = |v_k|^2$ which is a dimensionless quantity with $E=\hbar^2k^2/(2\bar{m})$. At zero temperature, 
the total number density $N = \frac{k_F^3}{6\pi^2}$ is related to $n(E)$ by $N = \int n(E) d^3k/(2\pi)^3$. Thus,   
to calculate the chemical potential $\mu$, we make use of the following equation 
\bea
\frac{k_F^3}{6\pi^2}=\frac{1}{2}\int\left(1-\frac{\epsilon_k-\mu}{\sqrt{\Delta_k^2+(\epsilon_k-\mu)^2}}\right)\frac{d^3k}{(2\pi)^3}
\label{eq:N}
\eea
\begin{figure}[t]
\centering
 \includegraphics[height=1.7in, width=5.4in]{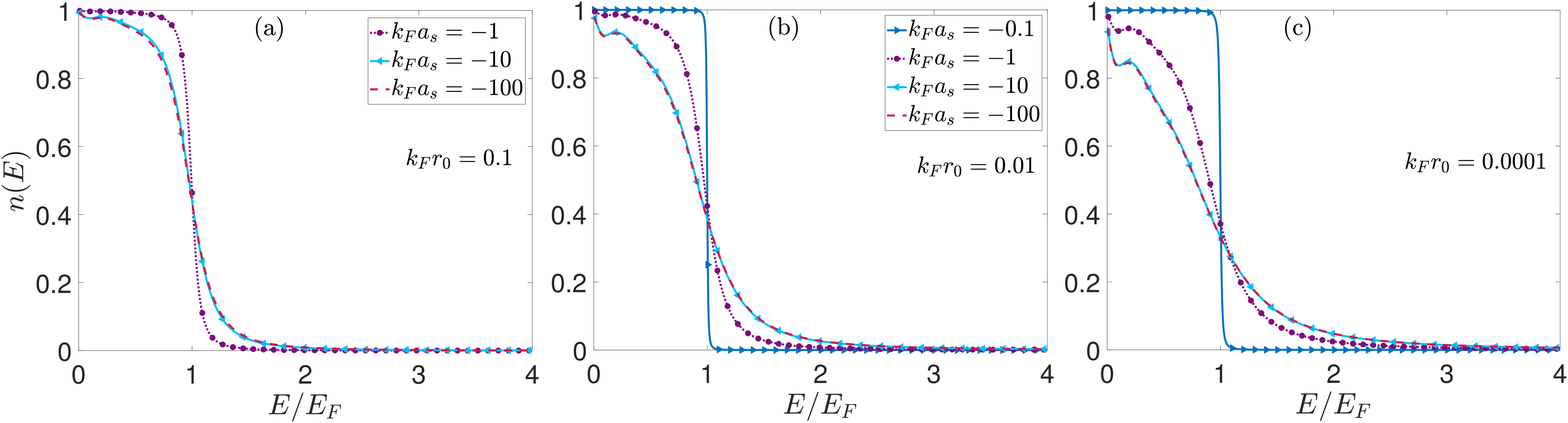}
 \centering
 \caption{The variation of superfluid density with energy $E$ (in unit of $E_F$) for the same set of parameters as in figure (\ref{fig.1}). The values of $a_s$ in (c) are same as in (b).}
 \label{fig.2}
\end{figure}
\begin{figure}[t]
\centering
\includegraphics[height=1.9in, width=4.8in]{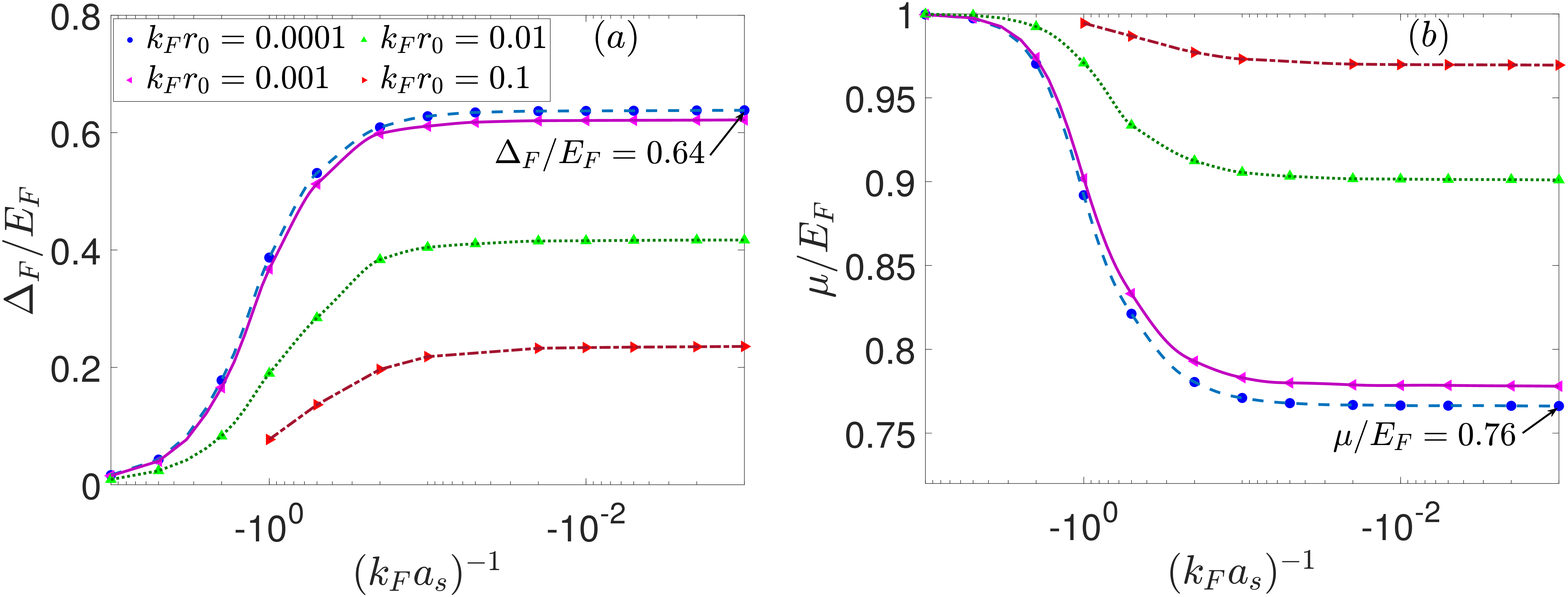}
 \centering
 \caption{Superfluid gap $\Delta_F$ at $k=k_F$ (a) and chemical potential $\mu$ (b) in Fermi energy scale are plotted as a function of inverse scattering length (in unit of $k_F$) for different ranges of interaction.}
 \label{fig.3}
\end{figure}

\section{Results and discussions}\label{sec.3}
The wave number- or energy-dependence of superfluid gap for different values of $a_s$ and three values of $r_0$, namely, $r_0 = 0.1 k_F^{-1}$, $r_0 = 0.01 k_F^{-1}$ and $r_0 = 0.0001 k_F^{-1}$ is shown in the subplots (a), (b) and (c) of  figure (\ref{fig.1}), respectively. We numerically solve the two coupled equations (\ref{eq:delk}) and (\ref{eq:N}). We resort to an iterative procedure to obtain numerically the convergent values of $\Delta_k$ and $\mu$. We notice that $\Delta(k) \equiv \Delta_k$ as a function of $k$ has the peak value at $k=0$ and monotonically decreases as $k$ increases. Higher the value of $a_s$, higher is $\Delta(k)$. With decreasing $r_0$,  $\Delta(k)$ increases and saturates as $r_0 \rightarrow 0$. The figure (\ref{fig.1}) further shows that, as  $r_0$  decreases below $0.01$ k$_F^{-1}$, the results for $\Delta(k)$ in the limit $a_s \rightarrow - \infty$ become convergent. $\Delta(0)$ for $a_s \rightarrow - \infty$ is found to be close to  the $k$-independent value $\Delta = 0.68E_F$ calculated using the zero-range pseudo-potential in the unitary limit \cite{Deb_JPB:2005}.    
\begin{figure}[htbp]
\centering
\includegraphics[height=1.9in, width=4.0in]{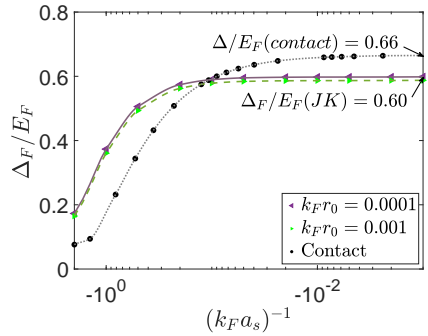} 
 \centering
 \caption{Comparison between the gaps ($\Delta_F$) as a function of $(k_F a_s)^{-1}$ for JK and contact potentials. The values of $\Delta_F$ for JK potential with $r_0=0.0001k_F^{-1}$ and that for contact potential in the unitary limit are found to be $0.60E_F$ and $0.66E_F$, respectively.}
 \label{fig.4}
\end{figure}

The energy-dependence of superfluid density $n(E)$  for different values of $a_s$ and the three values of $r_0$  is illustrated  in figure (\ref{fig.2}). For small values of $a_s$ in the zero-range limit, $n(E)$ resembles  that of standard Fermi-Dirac distribution which then deviates as $|a_s|$ increases or range becomes finite.    

The superfluid gap $\Delta_F = \Delta(k_F)$ at Fermi energy  and chemical potential $\mu$ in unit of $E_F$ are plotted as a function of dimensionless interaction strength $(k_Fa_s)^{-1}$ in figure (\ref{fig.3}) for different values of $r_0$. The variation of $\Delta_F$ as a function of $(k_F a_s)^{-1}$ in figure (\ref{fig.3}a) shows qualitatively similar behavior as in the contact interaction case \cite{Deb_JPB:2005}, and in the zero range limit its value is close to that of the contact case. Figure (\ref{fig.3}b) shows that, as $|k_F a_s|$ increases $\mu$ decreases. In the zero range limit of JK potential, the variation of $\mu$ with $(k_F a_s)^{-1}$ closely resembles to that for the contact interaction as given in Ref.\cite{Deb_JPB:2005}. In figure (\ref{fig.4}) we show the comparison between the variations of gap for the zero-range limit of the JK potential and the contact potential as a function $(k_Fa_s)^{-1}$. Qualitatively they are similar. In the unitarity limit their quantitative values which are found to be $0.66E_F$ and $0.60E_F$ for contact and JK potential, respectively, agree quite well.  

\begin{figure}[t]
\centering
\includegraphics[height=1.9in, width=4.0in]{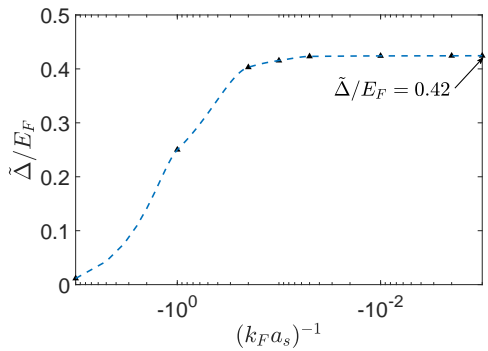} 
 \centering
 \caption{ $\tilde\Delta$ as a function of $(k_F a_s)^{-1}$ for $k_Fr_0 = 0.0001$.}
 \label{fig.5}
\end{figure}

 Figure (\ref{fig.1}) shows that the wave number dependence of $\Delta$ for finite-ranged JK potential or even in the zero range limit of this potential is quite significant. So, the question arises whether it is appropriate to compare $\Delta_F$ in the limit $r_0 \rightarrow 0$ with the energy-independent gap for the contact interaction. A more reliable quantity for comparison may be a mean gap  averaged over the full width at half maximum (FWHF) of the gap function. We define this energy-averaged mean gap by  
 \bea 
 \tilde\Delta = \frac{1}{\xi} \int_0^{\xi} \Delta(E) d E
 \label{eq.19}
 \eea 
where $\xi$ is the energy corresponding to the FWHM of the $\Delta(k)$. In figure (\ref{fig.5}) we have plotted $\tilde{\Delta}$  as a function of $(k_F a_s)^{-1}$ for $r_0 = 0.0001 k_F^{-1}$.  In the limit $a_s \rightarrow -\infty$, $\tilde{\Delta} = 0.42 E_F$.

 Most of the theoretical and experimental works on the superfluid gap of a unitary gas report the values of energy-independent gap for different system parameters. The recent experiment with superfluid $^6$Li using momentum and specially resolved radio frequency spectroscopy \cite{Hoinka_Nat_Phy:2017} reports  $\Delta=(0.39\pm0.03)E_F$  in the unitarity regime. There are numerous approaches to calculate the paring gap in the unitarity regime. Most of the calculations involve the quantum Monte-Carlo (QMC) technique. The first  QMC calculation was done with up to 40 particles with a modified Poschl-Teller interaction potential with $k_Fr_0\approx0.3$ \cite{Carlson_PRL:2003}. The estimated gap was $\Delta\sim0.54E_F$. 
The inclusion of polarization correction provides a better result $\Delta\approx0.49E_F$. Diffusion Monte-Carlo calculations with improved trial function yield  $\Delta=0.45E_F$ \cite{Carlson_PRL:2008} for larger number of particles with a better extrapolation to $k_Fr_0\rightarrow0$. A detailed comparison between the strong coupling theory and experimental results is made  in Ref. \cite{Ohashi_PRA:2018}. The QMC results are in reasonable agreement with the experimental ones \cite{Hoinka_NatPhy:2017} for $r_0 = 0$. In our calculations, we numerically evaluate the gap for the entire range of energy. For the sake of comparison to other works, we choose the gap $\Delta_F$, that is $\Delta$ at $E=E_F$ and the energy-averaged gap $\tilde{\Delta}$ as defined in equation (\ref{eq.19}).  We find that 
$\Delta_F = 0.64  E_F$   in the limit $k_Fr_0\rightarrow0$ as displayed in figure (\ref{fig.3}a). Though this value is substantially higher than that of QMC prediction, it is slightly lower than that ($\Delta = 0.69 E_F$) predicted by mean-field BCS-Leggett theory \cite{BCS-Leggett}. However, our study shows that $\tilde{\Delta}$ as plotted in figure (\ref{fig.5}) agrees quite well with that predicted by the recent QMC calculations as well as with the experimental value. We have calculated the equation-of-states (EoS) which is given as $\bar{E} = \zeta\bar{E}_{FG}$ with $\bar{E}_{FG}$ is the ground state energy per particle of non-interacting Fermi gas and $\bar{E}$ being the energy par particle for the interacting system. We find that the Bertsch parameter $\zeta = 0.69$ in the zero-range limit of the unitarity regime which has good agreement with the Monte Carlo simulation results \cite{Carlson_PRL:2003}.    

\section{Conclusions}\label{sec.4}

In conclusion, we have demonstrated that the finite  range of interaction leads to quite significant energy-dependence in the pairing  gap and superfluid density both in the non-unitary and unitary regimes. The gap as a function of energy for  a fixed scattering length and a fixed range is maximum at zero energy, and decreases monotonically as energy increases. For a small scattering length in the zero range limit, the energy-dependence of superfluid density resembles a step function like a standard Fermi distribution, for large scattering lengths and any value of range the density function deviates significantly from the step-like shape. The FWHM of density function is the Fermi energy. We have shown that our results on the gap and density functions approach convergent values as the range decreases below $10^{-2} k_F$.  For the sake of comparison of  our results with those for zero-range contact interaction, we have resorted in two different ways. First, we have considered  the value of the gap  at the  Fermi energy in the zero range limit. This value is found be close to the corresponding value obtained by BCS mean-field method using contact potential. Next, we have considered the mean gap averaged over energy within FWHM of the gap in the unitary limit. This mean gap is found to be close to the results obtained by other workers using quantum Monte Carlo simulation. From a theoretical point of view, we have made use of the exact two-body scattering solution of the JK potential to obtain the energy-dependent ${\mathbf T}$-matrix element that is employed in our calculations. Unlike the case of contact interaction, the use of this exact ${\mathbf T}$-matrix element allows us to calculate accurately the energy dependence of the gap without requiring any renormalisation of the interaction.  The effective range effects described in this paper may be experimentally realisable with a relatively dense cloud of ultracold fermionic atoms near a  Feshbach resonance  for which the effective range is finite, positive and likely to be tunable with an external field \cite{Dyke_PRA:2013, Mal_JPB1:2019}. In fact, the effective range near a narrow Feshbach resonance may become large and even negative \cite{Blackley_PRA:2014,Hazlett_PRL:2012,Dyke_PRA:2013}. In our study, we have not considered negative effective range which may also affect the pairing gap significantly \cite{Hu_PRA:2020}.  However, in the known neutral $s$-wave superfluids such as neutron stars,  negative effective range is unlikely to arise. Our work may serve as a precursor  towards  exploring  quantum simulation of nuclear superfluidity using cold atoms. 

Since the model potential we have used in our work, namely JK potential, is derived by inverse scattering method based on effective range expansion, the results presented here apply to all realistic interaction potentials that admit effective range expansion. There are two classes of JK potentials - one for negative scattering length which effectively describes attractive interaction and the other for positive scattering length to account for repulsive interaction. In this paper we have studied only the BCS side of the BCS-BEC crossover including the unitary regime and hence used only former  type of JK potentials. While the JK potential for negative scattering length is a two-parameter potential, that corresponding to the positive scattering length is  a three-parameter potential - the additional parameter being the binding energy of the last bound state of the actual potential. Therefore, in general the two potentials do not smoothly match in the unitarity limit, that is $a_s\rightarrow\pm\infty$. Only if one assumes that the binding energy is given by $\hbar^2/(2\bar{m}a_s^2)$ ($\bar{m}$ is the reduced mass), then the two potentials reduce to the same form in the unitarity limit \cite{Deb_IJMPB:2016}, and so under this assumption it may be possible to explore the smooth BCS-BEC crossover with JK potentials. This also suggests that in general the BCS-BEC crossover may not be  continuous or smooth when the energy of the last bound state does not go to zero in the limit $a_s\rightarrow\infty$. We hope to address these issues in our future communication. One of the most salient features of these potentials is that in the unitarity limit they take the form of Posch-Teller potentials which admit exact analytical solutions in one dimension. Since the partial-wave Sch\"{o}dinger equation is an effective one dimensional equation in one side, one can use the first odd solution of the Posch-Teller equation as the exact finite-ranged $s$-wave solution in the unitary limit within the framework of effective range expansion. So, it is expected that the use of JK potentials will enable one to study the physics in the unitarity regime more accurately.

\section*{Acknowledgment} 
SM is thankful to Council of Scientific and Industrial Research (CSIR), Govt. of India, for a support.

% \begin{thebibliography}{100}
%  \bibitem{Cooper_PR_1956} L. N. Cooper, {Phys. Rev. {\bf104}(4), 1189–1190 (1956)}.
%  \bibitem{Grasso_PRA:2003}M. Grasso and M. Urban, {Phys. Rev. A {\bf68}, 033610 (2003)}.
% \end{thebibliography}
%\section*{References}
\bibliographystyle{ieeetr}
\bibliography{Finite_Range_revised2}

\end{document}